\documentstyle[aps,twocolumn,epsf]{revtex}
\begin{document}
\title{Phase measurements with weak reference pulses}
\author{S.J. van Enk\\
Bell Labs, Lucent Technologies,
Room 2C-401\\
600-700 Mountain Ave,
Murray Hill NJ 07974}
\maketitle
\begin{abstract}Quantum state discrimination for two coherent states with opposite phases as measured relative to a reference pulse is analyzed as functions of the intensities of both the signal states and of the reference pulse. This problem is relevant for Quantum Key Distribution with phase encoding. We consider both the optimum measurements and simple measurements that require only beamsplitters and photodetectors.\end{abstract}
\medskip
\section{Introduction}
Suppose we are given a light pulse in one of two possible coherent states $|\alpha\rangle$ or $|-\alpha\rangle$ (with $\alpha$ real and positive) and we are to guess which one of the two we have in our possession. The minimum error probability is \cite{helstrom} 
\begin{equation}\label{min}
P_{{\rm min}}=\frac{1-\sqrt{1-|\langle \alpha|-\alpha\rangle |^2}}{2}
=\frac{1-\sqrt{1-\exp(-4\alpha^2)}}{2}.
\end{equation}
A measurement that achieves this minimum error probability using linear optics, photon counters and feedback, but which is hard to implement, was given by Dolinar in \cite{dolinar}. 
An alternative optimum scheme, not requiring feedback, but still complicated, was considered in \cite{sasaki} (see also \cite{osaki}).
A much simpler ``near-optimum'' scheme using linear optics, photon counters but no feedback was presented by Kennedy in \cite{kennedy}, which achieves an error probability of
\begin{equation}\label{Ken}
P_{{\rm Ken}}=\frac{\exp(-4\alpha^2)}{2}.
\end{equation}
This scheme is near optimal in the sense that in the limit of $\alpha\rightarrow\infty$ (the relevant limit for {\em classical} communication) $P_{{\rm Ken}}\rightarrow 2P_{{\rm min}}$.

The standard technique of homodyne detection \cite{yuen} would lead to an error probability (for more details, see below)
\begin{equation}\label{hom}
P_{{\rm hom}}=\frac{1}{\sqrt{2\pi}}\int_{2\alpha}^{\infty}\! {\rm d}\tau \exp(-\tau^2/2),
\end{equation}
which is clearly inferior to the Kennedy measurement for large amplitudes, but which is superior (although not optimal)
for small amplitudes $\alpha$, which is the relevant limit for {\em quantum} communication.
In particular, experimental implementations of Quantum Key Distribution protocols (see for example \cite{QKD}),
rely on the use of weak laser pulses with average numbers of photons of $\alpha^2\approx 0.1-0.2$ with the express goal of producing nonorthogonal and hence not perfectly distinguishable quantum states. 
 
Now lasers actually do not produce coherent states but mixtures of coherent states described by density matrices of the form\cite{moelmer}
\begin{equation}
\rho_{|\alpha|}=\int \frac{{\rm d}\phi}{2\pi} |\alpha e^{i\phi}\rangle \langle \alpha e^{i\phi}|.
\end{equation}
The ``phase'' of a laser field can only be defined relative to another laser beam. The problem of distinguishing two phases $\phi_0=0$ and $\phi_1=\pi$ of a faint laser beam with amplitude $\alpha$ in the presence of a reference pulse with amplitude $\beta$ can be formulated as distinguishing two mixed states $\rho_0$ and $\rho_1$ given by\cite{enkfuchs}
\begin{eqnarray}\label{ba}
\rho_k=\int \frac{{\rm d}\phi}{2\pi} |\beta e^{i\phi}\rangle \langle \beta e^{i\phi}|
\otimes |\alpha e^{i(\phi+\phi_k)}\rangle \langle \alpha e^{i(\phi+\phi_k)}|,
\end{eqnarray}
for $k=0,1$.
The detection schemes mentioned above assume that an absolute phase standard is present and indeed explicitly
require an in principle infinite amount of auxiliary light with a known phase (i.e., $\beta\rightarrow\infty$).
But suppose one has at one's disposal only a phase reference pulse of finite amplitude $\beta$. How well can one distinguish the two phases given this restricted resource?
This problem shows up in QKD where phase is used to encode information. 
In such a case a reference pulse is sent along with the signal pulse, typically over the same fiber. But the reference pulse may not be chosen arbitrarily strong as some of that light may cross over to and thus contaminate the signal.
Interestingly, even when polarization is the degree of freedom encoding the information the light pulses are properly described by states of the form (\ref{ba}) with $\beta=\alpha$ \cite{norbert}.

In Section \ref{LO} we consider simple measurements that require linear optics and photon counters but no feedback. We generalize Kennedy's measurement and homodyne detection to setups with finite reference pulses. We also construct a whole class of measurements that includes those two measurements and other improved measurements.
In Section \ref{OPT} we consider the optimum measurement for quantum state discrimination of the states (\ref{ba}).
\section{Measurements with linear optics}\label{LO}
\subsection{Generalized Kennedy measurement}
\begin{figure}\leavevmode
\epsfxsize=8cm \epsfbox{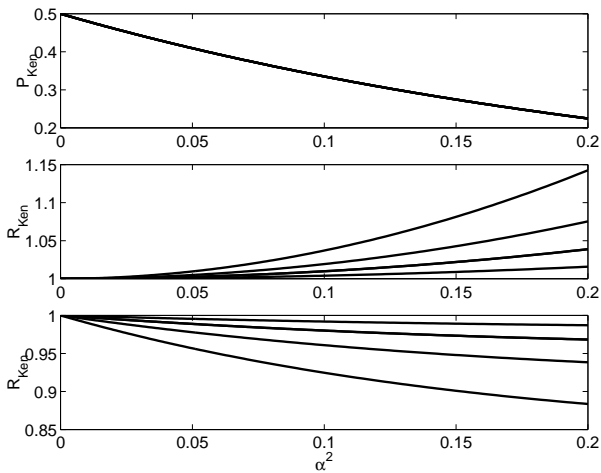} \caption{The upper graph gives the probability $P_{{\rm KEN}}$ as a function of the number of photons $\alpha^2$ in the unknown state; The middle and bottom graphs give the ratios of the error probabilities
$\tilde{P}_{{\rm KEN}}/P_{{\rm KEN}}$ and of the distinguishability measures
$\tilde{D}_{{\rm KEN}}/D_{{\rm KEN}}$, respectively, 
as functions of $\alpha^2$ for different values of the number of photons in the reference pulse, 
$\beta^2=1,2,4,10$, where increasing values of $\beta$ correspond to increasing distinguishability and decreasing error probability.}
\label{PKEN1} \end{figure}

Kennedy's measurement combines the unknown state $|\pm \alpha\rangle$ on a beamsplitter with a  reference beam with amplitude $\beta=r\alpha/t$, where $r$ and $t$ are the absolute values of the reflection and transmission coefficients of the beamsplitter. In the limit $t\rightarrow 0$ one of the output ports will either have a coherent state of amplitude $2\alpha$ or the vacuum, depending on the phase of the unknown coherent state (the other output is useless).
If a photon is detected in that port, one is certain to have the state $|-\alpha\rangle$, if no photon is registered one guesses that the unknown state is $|\alpha\rangle$. The probability of a wrong guess is then given by (\ref{Ken}).
The limit of $t\rightarrow 0$, however, implies $\beta\rightarrow\infty$. Given a finite amount of light, we can generalize the Kennedy measurement by requiring that in the useful output port the amplitudes cancel if the state is $|\alpha\rangle$. This requires we keep
the relation $\beta=r\alpha/t$. One output port will then either contain the vacuum or a coherent state with amplitude $2r\alpha$. The corresponding error probability is then
\begin{equation}\label{Ken2}
\tilde{P}_{{\rm Ken}}=\frac{\exp(-4r^2\alpha^2)}{2}.
\end{equation}
For consistency the reflection coefficient for finite $\beta$ has to be chosen as
\begin{equation}\label{t}
r^2=\frac{\beta^2}{\alpha^2+\beta^2}.
\end{equation}
The error probability reduces to (\ref{Ken}) in the limit $\beta\rightarrow\infty$ and reduces to 1/2 for $\beta\rightarrow 0$, as it should. Viewed as a function of $\alpha$ and $\beta$, (\ref{Ken2}) is symmetric in its arguments because the measurement procedure is symmetric in the signal and reference states: the phase of one is defined only relative to the other. For the small values of $\alpha$ we are interested in, all error probabilities are close to 1/2.
A better measure to compare different probabilities may be generically defined as 
\begin{equation}
D=1-2P,
\end{equation}
in terms of the corresponding error probability. $D$ 
may be interpreted as a measure of distinguishability and ranges between 0 for identical states and 1 for orthogonal states. See Fig.~\ref{PKEN1} for plots of both the error probabilities and the corresponding measure of distinguishability for the (generalized) Kennedy measurement.
If we consider signal states with $\alpha^2=0.1$ 
the Figure shows that 10 photons in the reference pulse are sufficient to be within 1\% of the best achievable (for either measure) for the Kennedy measurement, and even just 1 photon brings one within 7\%. 
\subsection{Generalized homodyne detection}
\begin{figure}\leavevmode
\epsfxsize=8cm \epsfbox{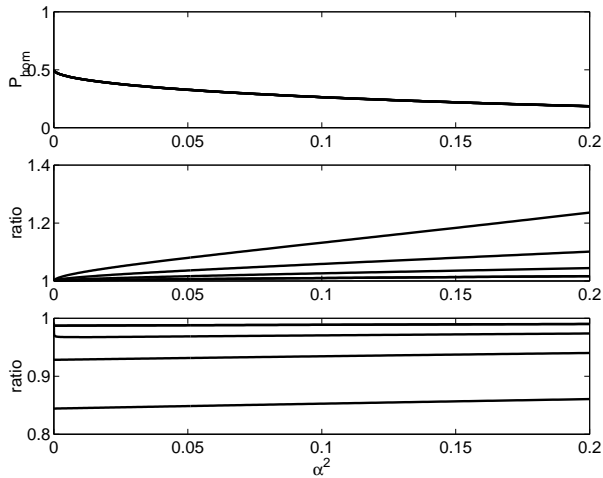} \caption{Same as Figure~\ref{PKEN1} but for homodyne measurements.}
\label{Phom1} \end{figure}

In a homodyne detection scheme one splits the unknown coherent state on a 50/50 beamsplitter with 
a reference light beam of amplitude $\beta$ and measures the difference in photon number between the two output ports. The two output modes are in coherent states with amplitudes $(\beta\pm \alpha) /\sqrt{2}$.
In the limit of $\beta\rightarrow\infty$ the difference between the expected photon numbers $|\beta\pm \alpha|^2/2$ becomes linear in $\alpha$ and thus homodyne detection directly measures the amplitude.
For finite $\beta$ we use a similar strategy: we guess that the output port with the larger number of detected photons is associated with an amplitude $(\beta+\alpha)/\sqrt{2}$ (in the case of an equal number of photons we make a random guess).
The probability to detect $n$ photons in a coherent state with amplitude $(\beta\pm\alpha)/\sqrt{2}$ is 
\begin{equation}
P_\pm(n)=\exp(-N_\pm)\frac{N_\pm^n}{n!},
\end{equation}
in terms of the expected number of photons $N_\pm=|\beta\pm\alpha|^2/2$.
Our procedure gives a wrong result if the larger coherent state is found to contain fewer photons than the smaller one. Moreover, if we find an equal number of photons we will have to make a random guess. The total error probability, therefore, is
\begin{eqnarray}\label{Pd}
\tilde{P}_{{\rm hom}}&=&\sum_{n=0}^\infty \sum_{m=n+1}^\infty P_+(n)P_-(m)
+\frac{1}{2}\sum_{n=0}^\infty P_+(n)P_-(n).
\end{eqnarray}
Just as for the generalized Kennedy measurement, this probability function is symmetric in $\alpha$ and $\beta$. It is plotted in Fig.~\ref{Phom1}.
Just as for the Kennedy measurement a reference pulse containing 10 photons is sufficient to reach a distinguishability of 99\% of the best a homodyne measurement can achieve, while a reference pulse with just 1 photon on average brings one to within about 15\% at $\alpha^2=0.1$.

For completeness we note that in the limit $\beta\rightarrow\infty$ both probability distributions $P_+(n)$ and $P_-(n)$ approach Gaussian distributions. That is, defining continuous variables $x_{\pm}=(n-N_\pm)/\sqrt{2N_\pm}$
we get
\begin{equation}
P_{\pm}(x_{\pm})\rightarrow \frac{1}{\sqrt{2\pi}}\exp(-x_\pm^2),
\end{equation}
with the variable $x_{\pm}$ ranging from $-\infty$ to $\infty$.
The probability (\ref{Pd}) reduces then to
\begin{eqnarray}\label{Pc}
\tilde{P}_{{\rm hom}}&\rightarrow& \frac{1}{2\pi}\int_{-\infty}^\infty {\rm d}x_-
\exp(-x_-^2)
\int_{x_--2\alpha}^\infty {\rm d}x_+ \exp(-x_+^2)\nonumber\\
&=&\frac{1}{\sqrt{2\pi}}\int_{2\alpha}^\infty {\rm d}x_+ \exp(-x_+^2/2),
\end{eqnarray}
which confirms (\ref{hom}). 
\subsection{Class of generalized measurements}
The generalized Kennedy and homodyne measurements are special cases of a whole class of similarly straightforward measurements. We can take {\em any} beamsplitter with arbitrary transmission and reflection coefficients whose absolute values $t,r$ can be parametrized without loss of generality as
\begin{equation}
r=\cos(\phi);\,\,t=\sin(\phi)\,\,0\leq \phi\leq\pi/4.
\end{equation}
 The expected numbers of photons in the two output ports, 
\begin{eqnarray}
N^{(1)}_{\pm}&=&\big|r\beta\pm t\alpha\big|^2\nonumber\\
N^{(2)}_{\pm}&=&\big|t\beta\mp r\alpha\big|^2,
\end{eqnarray}
respectively, depend on the phase of the unknown state. If we found $n$ photons in the first detector, $m$ in the second, we calculate the joint probabilities
\begin{equation}
P_{\pm}(n,m)=P^{(1)}_{\pm}(n)P^{(2)}_{\pm}(m),
\end{equation}
where 
\begin{equation}
P^{(k)}_{\pm}(n)=
\exp(-N_\pm^{(k)} )\frac{\left(N_\pm^{(k)}\right)^n}{n!}\,\,\,(k=1,2)
\end{equation}
If $P_+(n,m)>P_-(n,m)$ we guess  that we had the state $|\alpha\rangle$, 
if $P_+(n,m)<P_-(n,m)$ we guess that we had the state $|-\alpha\rangle$, if the two probabilities happen to be equal we make a random guess. This corresponds to maximizing the {\em conditional} probabilities for the unknown state to be $|\pm \alpha\rangle$ given $n$ and $m$ clicks in the respective photodetectors. The error probability is
\begin{eqnarray}
\tilde{P}(\phi)&=&\frac{1}{2}\sum_{P_->P_+} P_{+}(n,m)+\frac{1}{2}\sum_{P_+>P_-} P_{-}(n,m)
\nonumber\\&&+\frac{1}{2}\sum_{P_+=P_-}P_+(n,m),
\end{eqnarray} 
where the first summation runs over all pairs $n,m$ such that $P_-(n,m)>P_+(n,m)$, etcetera.
The generalized homodyne and Kennedy measurements are special cases of this general measurement for angles $\phi_{{\rm hom}}=\pi/4$ and $\phi_{{\rm Ken}}=\arctan (\alpha/\beta)$.

\begin{figure}\leavevmode
\epsfxsize=8cm \epsfbox{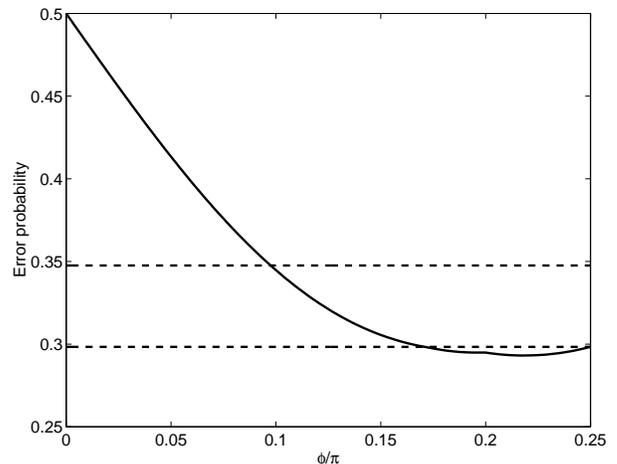} \caption{$\tilde{P}(\phi)$ as a function of $\phi/\pi$ (solid curve) for $\alpha^2=0.1$ and $\beta^2=1$. The dashed lines give the values of $\tilde{P}_{{\rm KEN}}$ (around 0.35) and $\tilde{P}_{{\rm hom}}$ (around 0.30) as reference. 
}
\label{phi1} \end{figure}

\begin{figure}\leavevmode
\epsfxsize=8cm \epsfbox{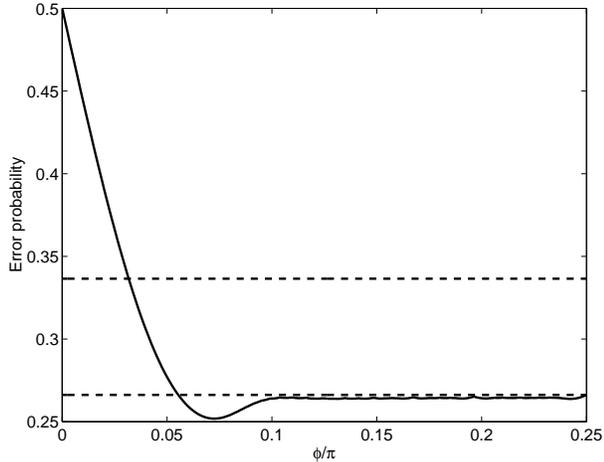} \caption{$\tilde{P}(\phi)$ as a function of $\phi/\pi$ (solid curve) for $\alpha^2=0.1$ and $\beta^2=10$. The dashed lines give the values of $\tilde{P}_{{\rm KEN}}$ and $\tilde{P}_{{\rm hom}}$ as reference. The value of $\beta$ is sufficiently large for  $\tilde{P}(\phi)$ to improve upon standard homodyne detection, since $P_{{\rm hom}}\approx 0.26$ and the minimum for $\tilde{P}(\phi)$ is 0.25. Also note that $\tilde{P}(\phi)$ starts to approach $P_{{\rm hom}}$ more and more (when compared to the previous plot) especially for $\phi/\pi>0.1$.
}
\label{phi10} \end{figure}

In Figures \ref{phi1} and \ref{phi10} we plot the error probability $\tilde{P}(\phi)$ as a function of $\phi$ for the case of $\alpha^2=0.1$ and $\beta^2=1$ and 10, respectively. 
For this small value of $\alpha$ homodyne detection is better than Kennedy's measurement and for certain values of $\phi$ the error probability $\tilde{P}(\phi)$ is even smaller.
For $rt\neq 0$ the probability distributions again approach Gaussian distribution functions
in the limit $\beta\rightarrow\infty$. Remarkably, $\tilde{P}(\phi)$ approaches $P_{{\rm hom}}$ for any nonzero $\phi$ in that same limit, but the limit is not uniform. Although an arbitrary beamsplitter with nonzero reflection and transmission coefficients does not improve upon standard homodyne detection in the limit $\beta\rightarrow \infty$, for finite values of $\beta$ improvement is in fact possible as illustrated in Fig.~\ref{phi10}. 
\section{Optimum measurement}\label{OPT}
The measurements considered in the previous Section were chosen for their simplicity but do not allow one to achieve the minimum error probability. For the case of pure coherent states the minimum error probability was given in (\ref{min}), its generalization to mixed states is\cite{fuchs}
\begin{equation}\label{err}
P_{{\rm err}}=\frac{1}{2}-\frac{1}{4}{\rm Tr} |\rho_1-\rho_0|,
\end{equation}
which gives the minimum error probability for distinguishing two mixed states that are {\em a priori} equally likely. This expression reduces to (\ref{min}) for $\beta\rightarrow\infty$.
The density matrix $\rho_1-\rho_0$ can be written in the number-state basis as
\begin{eqnarray}
\rho_1-\rho_0&=&\int \frac{{\rm d}\phi}{2\pi} |\beta e^{i\phi}\rangle \langle \beta e^{i\phi}|
\nonumber\\
&&\otimes \big[|\alpha e^{i\phi}\rangle \langle \alpha e^{i\phi}|-
|-\alpha e^{i\phi}\rangle \langle -\alpha e^{i\phi}|\big]\nonumber\\
&=&e^{-\alpha^2-\beta^2}
\sum_{n,m,p,q}\delta(n-m+p-q)\big(1-(-1)^{p+q}\big)\nonumber\\
&&\frac{\beta^{n+m}\alpha^{p+q}}{\sqrt{n!m!p!q!}}|n\rangle\langle m|\otimes |p\rangle\langle q|.
\end{eqnarray}
Here we are mostly interested in the limit of small $\alpha$. The lowest-order term in an expansion in powers of $\alpha$ of the density matrix $\rho_1-\rho_0$ is linear in $\alpha$ and contains two terms:
\begin{eqnarray}
\rho_1-\rho_0&\approx&
e^{-\beta^2}\sum_{n\geq 0}\frac{2\beta^{2n+1}\alpha}{\sqrt{n!(n+1)!}}\nonumber\\
&&|n\rangle\langle n+1|\otimes |1\rangle\langle 0|+
|n+1\rangle\langle n|\otimes |0\rangle\langle 1|.
\end{eqnarray}
Its eigenvectors are then of the form 
\begin{equation}\label{eig}
|\psi_n^{\pm}\rangle=\frac{|n\rangle\otimes|1\rangle\pm |n+1\rangle\otimes|0\rangle}{\sqrt{2}},
\end{equation}
with eigenvalues
\begin{equation}
\lambda_n^{\pm}=\pm\frac{2\beta^{2n+1}\alpha
e^{-\beta^2}
}{\sqrt{n!(n+1)!}},\end{equation}
where $n\geq 0$ an integer. 
The optimum measurement achieving the minimum error probability is then a projective measurement onto the eigenstates (\ref{eig}).
It is an open question how this measurement can be implemented.  
In the limit that $\alpha=\beta\rightarrow 0$ the optimum measurement is equivalent to the generalized Kennedy and homodyne detection schemes.

\begin{figure}\leavevmode
\epsfxsize=8cm \epsfbox{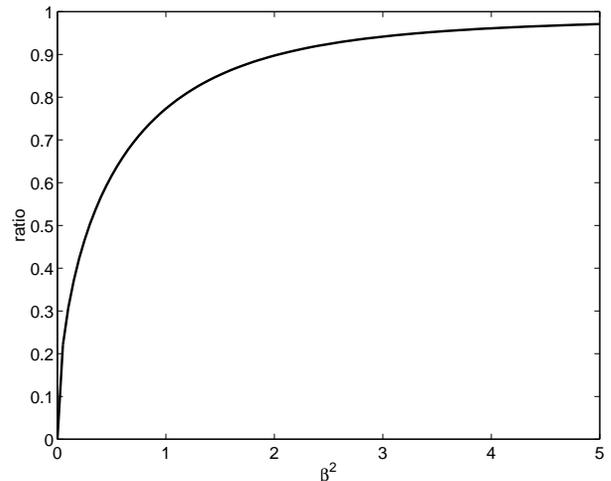} \caption{The ratio of the distinguishability measures $D_{{\rm err}}/D_{{\rm min}}$ as a function of the number of photons in the reference pulse $\beta^2$ in the limit of small signal amplitude $\alpha$.}
\label{Trrho} \end{figure}

We can now evaluate $D_{{\rm err}}=1-2P_{{\rm err}}$ as
\begin{equation}
D_{{\rm err}}=\frac{1}{2}\sum|\lambda_n^{\pm}|
\approx 2\alpha e^{-\beta^2}\sum_{n\geq 0}\frac{\beta^{2n+1}}{\sqrt{n!(n+1)!}}.
\end{equation}
In Fig.~\ref{Trrho} we plot the ratio $D_{{\rm err}}/D_{{\rm min}}$ as a function of $\beta^2$, where $D_{{\rm min}}=2\alpha$ corresponds to the limit of $\beta\rightarrow\infty$ and $\alpha$ small. 
\section{Conclusions}
We considered the question how well one can distinguish two faint laser pulses with opposite phases as a function of the intensity of the phase reference pulse. It turns out that even with a reference pulse containing just 1 photon on average one does reasonably well. For example, take 
a signal state that possesses 0.1 photons on average. Then,
depending on what measurement one considers, the distinguishability is  somewhere between 75\% and 95\% of the best achievable with an infinite reference pulse. We considered generalizations of Kennedy's measurement and homodyne detection, and the optimum measurement. 
\section*{Acknowledgements}
I thank Chris Fuchs for useful discussions.

\end{document}